\documentclass[epj,final]{svjour}
\usepackage{graphics}
\usepackage{amssymb}
\usepackage{amsmath}
\allowdisplaybreaks[2]
\begin{document}
\title{The Z=82 shell closure in neutron-deficient Pb isotopes}
\author{M. Bender\inst{1}
        \thanks{\emph{Present address:}
          Service de Physique Nucl{\'e}aire Th{\'e}ori\-que
          et Physique Math{\'e}matique, CP229,
          Universit{\'e} Libre de Bruxelles,
          B--1050 Bruxelles, Belgium.
        } \and
        T. Cornelius\inst{2}       \and
        G. A. Lalazissis\inst{3,4} \and
        J. A. Maruhn\inst{2,4}     \and
        W. Nazarewicz\inst{5-7}    \and
        P.--G. Reinhard\inst{4,8}
}
\institute{Gesellschaft f\"ur Schwerionenforschung,
           Planckstrasse 1, 
           D--64291 Darmstadt, Germany.
           \and
           Institut f\"ur Theoretische Physik,
           Universit\"at Frankfurt,
           Robert-Mayer-Strasse 8--10, 
           D--60325 Frankfurt am Main, Germany.
           \and
           Department of Theoretical Physics,
           Aristotle University of Thessaloniki,
           Gr--54006 Thessaloniki, Greece.
           \and
           Joint Institute for Heavy-Ion Research, 
           Oak Ridge National Laboratory,
           P. O. Box 2008, Oak Ridge, Tennessee 37831.
           \and 
           Department of Physics and Astronomy, 
           University of Tennessee, 
           Knoxville, Tennessee 37996.
           \and
           Institute of Theoretical Physics, 
           Warsaw University, 
           ul. Ho\.za 69, PL--00681, Warsaw, Poland.
           \and
           Physics Division, Oak Ridge National Laboratory,
           P. O. Box 2008, Oak Ridge, Tennessee 37831.
           \and
           Institut f\"ur Theoretische Physik II,
           Universit\"at Erlangen--N\"urnberg,
           Staudtstrasse 7, D--91058 Erlangen, Germany.
}
 \date{October 23 2001}
% 
%=======================================================================
%
\abstract{
Recent mass measurements show a substantial weakening of the binding
energy difference \mbox{$\delta_{2p} (Z,N) = E(Z-2,N) - 2 E(Z,N) +
E(Z+2,N)$} in the neutron-deficient Pb isotopes. As $\delta_{2p}$ is
often attributed to the size of the proton magic gap, it might be
speculated that reduction in $\delta_{2p}$ is related to a weakening
of the spherical \mbox{$Z=82$} shell. We demonstrate that the observed
trend is described quantitatively by self-consistent mean-field models
in terms of deformed ground states of Hg and Po isotopes.
}
\maketitle
%
%=======================================================================
%
\section{Introduction}
The weakening or ``quenching'' of spherical shell closures when going 
away from the valley of stability to weakly bound nuclei is a phenomenon 
of great current interest. It is now well established for neutron-rich 
\mbox{$N=20$} and \mbox{$N=28$} isotones.
The study of excited states of nuclei around $^{32}$Mg populated by
$\beta^-$-decay \cite{Gui84a} strongly indicates that neutron-rich
\mbox{$N=20$} nuclides are deformed. This is consistent with direct mass 
measurements which, at magic numbers 20 and 28, do not yield the familiar 
drop in the two-neutron separation energy \cite{Vie86a,Orr91a,Zho91a}.
Energies and  $B(E2)$ transition probabilities obtained from Coulomb excitation
of radioactive beams of nuclides around $^{44}$S demonstrate also that
\mbox{$N=28$} isotones below $^{48}$Ca show strong collectivity 
in contradiction with a spherical shell closure \cite{Gla97a}.
Also, there are experimental hints from $\beta^-$ decay studies of
$^{80}$Zn
\cite{Kra88aE} and the systematics of $2^+$ and $4^+$ excitation energies 
in Cd and Pd isotopes \cite{Kau00aE} that the \mbox{$N=50$} and 
\mbox{$N=82$} shells are weakened when going towards neutron-rich nuclei.
Shell-quenching in neutron-rich systems has far-reaching consequences 
for astrophysics as it influences the $r$-process path \cite{Pfe97a}. 

The actual sequence of magic numbers in neutron-rich nuclei is strongly 
influenced by an increasing diffuseness of the neutron density, the closeness 
of the particle continuum, and/or the changes in the spin-orbit splitting 
\cite{Dob94,Lal98a}. For light neutron-rich \mbox{$N=28$} isotones,
this leads to an increased collectivity and even strong stable
quadrupole deformation; see, e.g., refs. \cite{Wer94a,Lal99a,Per00a}.
(For more discussion of shell structure of neutron-rich magic and semi-magic
nuclei, the reader is referred to ref.~\cite{Rei99} where a more complete 
literature review is given.)

The experimental signatures of ``magicity'' could sometimes be contradicting.
Consider, e.g., the \mbox{$N=40$} sub-shell in $^{68}$Ni. While it is clearly 
visible in the systematics of excited states \cite{Bro95a,Gra00a}, no sign 
of a shell effect is found when looking at two-neutron separation energies. 
Theoretically, this apparent inconsistency  can be explained in terms of  
dynamic correlations beyond the mean field  \cite{Rei00a}. The calculated 
single-particle spectra in $^{68}$Ni indeed show a \mbox{$N=40$} shell. 
However, the sudden increase in collectivity in the adjacent isotopes
obscures the picture when looking at isotopic energy differences
because one now compares nuclei with different intrinsic structure.
This example demonstrates that the various signatures of shell closures
are not always equivalent.

All examples mentioned so far concern neutron shells. The situation
seems to be different for protons. For light nuclei there is no
indication that the proton shell closures fade away towards the proton
drip line. The analysis of recent large-scale mass measurements of
proton-rich nuclei around \mbox{$Z=82$} at GSI \cite{Rad00a} indicates
a weakening of the proton shell gap in neutron-deficient Pb isotopes.
This is found when looking at the binding-energy differences
\cite{Novpre} and $Q_\alpha$ values \cite{Tot99a}. It is the aim of 
this paper to analyze these data from a theoretical perspective using 
self-consistent mean-field models.
%
%========================================================================
%
\section{Signatures of shell closures}
The notion of a shell closure comes from a mean-field description
where one has full insight into the single-nucleon energies
$\epsilon_k$ as eigenvalues of the single-particle Hamiltonian. 
A shell closure is associated with a large gap in the spectrum 
of $\epsilon_k$. It has to be stressed that $\epsilon_k$ 
are not equivalent to experimental single-particle energies $S_k$ 
which are measured as one-nucleon separation energies. To calculate
those, one has to take into account residual interactions (such as pairing
and coupling to low-lying vibrational modes), as well as rearrangement 
effects and the core polarization due to the unpaired nucleon; see, 
e.g., refs.~\cite{Ber80a,Rut98a}.

Bunchiness of levels and the presence of gaps in the single-particle spectrum 
can be quantified in terms of the shell-correction energy \cite{Bra72}
\begin{equation}
E_{\rm shell}
= \sum_k \epsilon_k - 
  \int \! {\rm d} \epsilon \, \epsilon \, g(\epsilon)
  \; ,
\end{equation}
where $g(\epsilon)$ is a smoothed level density. $E_{\rm shell}$ is
large and negative for magic nuclei because there the discrete sum has
a particularly low value while the smoothed expression varies only
slowly with mass number. It is to be noted that $E_{\rm shell}$ is a measure
of the deviation of the actual level density at 
the Fermi energy from the smoothed level density. For very heavy
systems beyond lead, the large level density inhibits pronounced
 shell gaps, and, yet, a dilution of levels suffices to produce
shell stabilization without magic gaps; see ref.~\cite{Ben01a}.
Shell corrections discussed in this study were calculated 
using the Green's function approach of
ref.~\cite{Kru00a}. The advantage of this new method is  
the proper treatment of unbound states which appear close to
the Fermi level in weakly bound (proton-rich or neutron-rich) nuclei.

Neither  $\epsilon_k$ nor $E_{\rm shell}$ are directly measurable
quantities. A quantity which is accessible from mass systematics 
is the so-called two-proton shell gap
\begin{equation}
\label{eq:tpgap}
\delta_{2p} (Z,N) 
= E (Z-2,N) - 2 E (Z,N) + E (Z+2,N)
\; ,
\end{equation}
which is the approximation to the second derivative of the nuclear 
binding energy. Let us assume that the single-particle energies do 
not change within the three isotones considered in Eq.~(\ref{eq:tpgap}), 
and that the change in the total binding energy comes from the variation 
of occupations around the Fermi surface. In such a case, Koopman's 
theorem states that $\delta_{2p}$ should represent \emph{twice the
gap in the corresponding single-particle spectrum}. 
This requires, however, that no dramatic rearrangements happen
among the three nuclei involved in $\delta_{2p}$.
For many nuclei, such rearrangements are small and $\delta_{2p}$ is 
a good representation of a shell gap. But this interpretation 
does not hold in situations where adding or removing just two 
nucleons induces a substantial change of the mean field. In spite of 
this weakness, it is worth noting that binding-energy differences 
involving even-even nuclei only -- like $\delta_{2p}$ and, to some 
extent, $Q_\alpha$ -- provide a cleaner signal and spectral
properties than the binding energy expressions connecting to the
neighboring  odd-mass systems. The double difference $\delta_p$ with
odd-mass neighbors mixes shell gap, and possibly rearrangement effects,
with pairing effects, which surely imply a dramatic rearrangement of
the pairing field by blocking.

Other experimental signatures of magic numbers are excitation energies of 
vibrational states and the associated electric transition rates. The gap 
in the single-particle spectrum sets the scale for the lowest excitations. 
Peaks in the systematics of the lowest $2^+$ states in even-even nuclei 
reflect the stiffness of the potential energy surface which is large in 
closed-shell nuclei. This signature is somewhat masked by the residual 
interaction which shifts the excitation energy considerably below the 
lowest particle-hole (or two-quasiparticle) energy. Compared to
binding-energy differences, the data on vibrational states have the
advantage  that they do not mix information from different nuclei. On
the other hand,  calculations of collective excitations require the use of
techniques  which go beyond the mean-field approach; hence the
information on magic  gaps is very difficult to extract.
%
%=======================================================================
%
\section{Theoretical framework}
We investigate the stability of the \mbox{$Z=82$} shell in the framework
of the self-consistent mean-field theory, either non-relativistic, using
Skyrme interactions (SHF), or the relativistic mean-field (RMF) approach.
As earlier studies have shown that parameterizations of the models 
often differ when extrapolated to weakly bound systems, we choose a 
sample of representative parameterizations which all give a very  
satisfactory description of stable nuclei but differ in details. Namely, 
we used the Skyrme interactions SkP \cite{SkP}, SLy6 \cite{SLyx}, 
SkI3, and SkI4 \cite{SkIx}. In the relativistic calculations, we employed 
NL3 \cite{NL3} and NL-Z2 \cite{Ben99a} parameterizations. The force 
SkP has effective mass \mbox{$m^*/m = 1$} and was originally designed 
to describe the particle-hole and particle-particle channel of the 
effective interaction simultaneously. The forces SLy6, SkI3, and SkI4 
stem from recent fits which already include data on exotic nuclei. Both
SkP  and SLy6 use the standard spin-orbit interaction. The forces SkI3/4
employ  a spin-orbit force with modified isovector dependence. SkI3
contains a  fixed isovector part analogous to the RMF, whereas SkI4 is
adjusted allowing free variation of the isovector spin-orbit term. The
RMF force NL-Z2 is  fitted in the same way as SkI3 and SkI4 to a similar
set of observables. Pairing is treated within the BCS approximation using
a zero-range delta force with the strength adjusted for each mean-field
parameterization  as described as in ref.~\cite{Ben00a}.
\clearpage
%
%=========
%
\begin{figure}[t!]
\centerline{\includegraphics{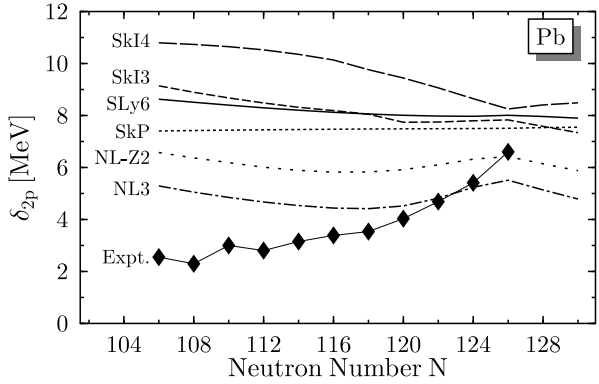}}
\caption{\label{fig:gap:sph}
Two-proton gap parameter $\delta_{2p}$ (\protect\ref{eq:tpgap})
for the chain of Pb isotopes obtained in several spherical mean-field 
models. Experimental values are marked with filled diamonds.
}
\end{figure}
%
%=========
%
%
%=========
%
\begin{figure}[b!]
\centerline{\includegraphics{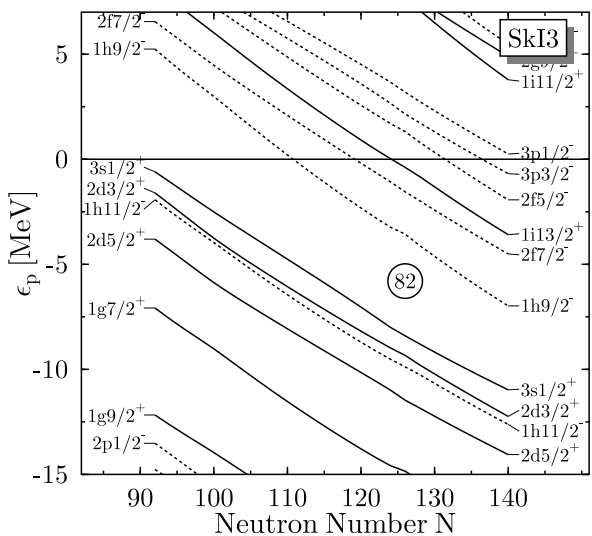}}
\caption{\label{fig:levels:sph:ski3}
Self-consistent single-proton energies $\epsilon_k$ for the 
chain of Pb isotopes calculated with  the spherical SkI3 model.
Other effective interactions give similar results. See 
ref.~ \protect\cite{Ben99a} for a comparison of calculated and
experimental spectra for $^{208}$Pb.
}
\end{figure}
%
%=========
%
%=======================================================================
%
\section{Results}
%
%--------------------------------------------------------------------------
%
\subsection{Spherical calculations}
We start with a presentation of purely spherical
calculations. By constraining the shape to be spherical, 
one can dramatically reduce polarization effects. Consequently, the 
assumptions behind the Koopman's theorem  can be met, and the clear 
correspondence between $\delta_{2p}$ and single-particle energies emerges.
Figure \ref{fig:gap:sph} shows the two-proton shell gap parameter  
$\delta_{2p}$ along the chain of Pb isotopes. The experimental data 
are compared with results of spherical mean-field calculations. While 
the experimental values decrease monotonously when going from $^{208}$Pb 
towards the proton-drip line, theoretical  results show a very different
trend.  Namely, $\delta_{2p}$ slightly \emph{decreases} with $Z$. The 
corresponding spherical single-proton energies in SkI3 are displayed in 
fig.~\ref{fig:levels:sph:ski3}. It is seen that the \mbox{$Z=82$} shell 
gap stays large  for all the Pb isotopes considered, and it grows slightly 
when approaching  the proton drip line. (Other parameterizations employed 
in our study  give similar results.) This result is consistent with 
fig.~\ref{fig:gap:sph}:  $\delta_{2p}$ reproduces twice the shell gap 
seen in the spectrum of $\epsilon_k$. 
This demonstrates that the idea beyond the two-nucleon shell gap as a
signature of shell closures is correct, provided the mean field does not 
change much between the nuclei  involved in its evaluation. The
restriction to spherical calculations enforces this feature.
%
%=========
%
\begin{figure}[t!]
\centerline{\includegraphics{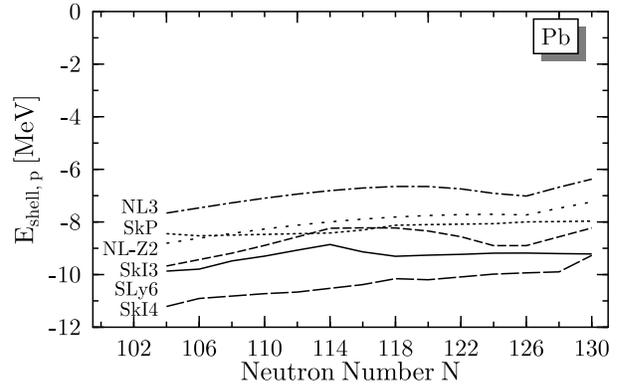}}
\caption{\label{fig:sc}
Spherical proton shell correction for the chain of Pb 
isotopes extracted from  self-consistent calculations.
}
\end{figure}
%
%=========
%

The stability of the \mbox{$Z=82$} shell is confirmed when inspecting
the proton shell correction energy $E_{\rm shell,p}$ extracted from
our self-consistent calculations; see fig.~\ref{fig:sc}.
There are some differences among the various forces as far as
the actual values are concerned. For instance, 
RMF models generally yield smaller shell correction than SHF models.  The
common feature seen in fig.~\ref{fig:sc} is that all the models 
predict the same trend, namely that the magnitude of
the proton shell correction is slightly
increasing when going proton-rich. One can thus say that all the 
indicators of shell effects considered in our study, i.e., 
single-particle energies, shell corrections, and  $\delta_{2p}$, give a
consistent picture of a stable \mbox{$Z=82$} shell for all 
neutron-deficient Pb isotopes. Contrary to what may be suggested by the
experimental plot of $\delta_{2p}$, the proton magic gap does not
decrease, but slightly increases when approaching the proton drip line.
% 
%=========
%
\begin{figure*}[ht!]
\centerline{\includegraphics{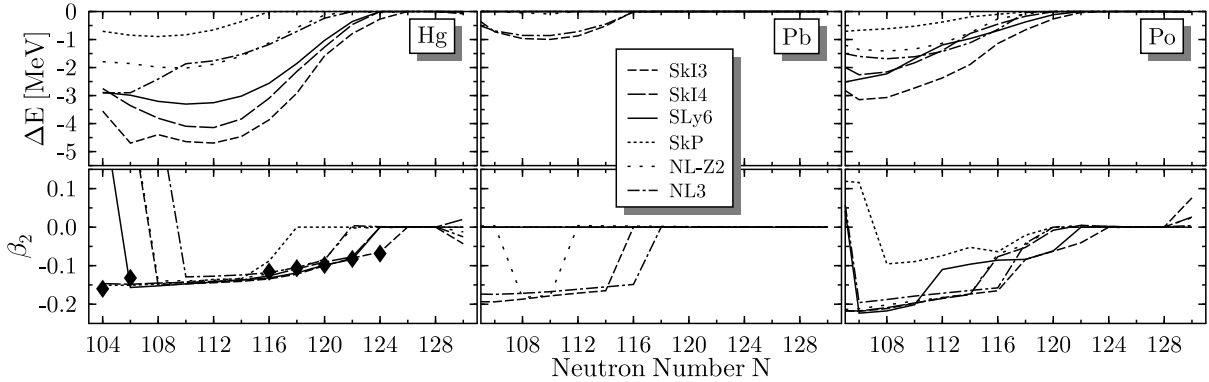}}
\caption{\label{fig:def}
Calculated deformation energy $\Delta E$ (top) and associated quadrupole 
deformation 
\mbox{$\beta_2 = 4 \pi \, \langle r^2 \, Y_{20} \rangle / (3 A r_0^2)$}
with \mbox{$r_0 = 1.2 \, A^{1/3} \; $fm} (bottom) for the ground states 
of even-even Hg, Pb, and Po isotopes. Experimental quadrupole deformations
(filled diamonds) are taken from ref.~\protect\cite{Ram01a}.
}
\end{figure*}
%
%=========
%
%------------------------------------------------------------------------
%
\subsection{Deformed calculations}
It is well known that shape coexistence is a common feature in 
neutron-deficient Hg, Pb, and   Po nuclei, and that most 
of the Hg and Po nuclei have deformed ground states \cite{Woo92,Jul01}.
Figure~\ref{fig:def} shows the calculated deformation energies $\Delta E$ 
(defined as a difference between  ground-state binding energies 
obtained in deformed and spherical calculations) and quadrupole 
deformations $\beta_2$ for even-even Hg, Pb, and   Po isotopic chains.
Most  nuclei have deformed ground states with an energy gain of 
several MeV. The potential energy surfaces calculated in this region
exhibit several competing minima, often leading to shape coexistence
and configuration mixing, see e.g., refs.~\cite{May77,Naz93a}.
In fact, the question if whether there were strongly deformed nuclei near
the  magic proton shell \mbox{$Z=82$} was already raised as early as 1972 
\cite{Fae72a,Cai73a}.
%
%=========
%
\begin{figure}[b!]
\centerline{\includegraphics{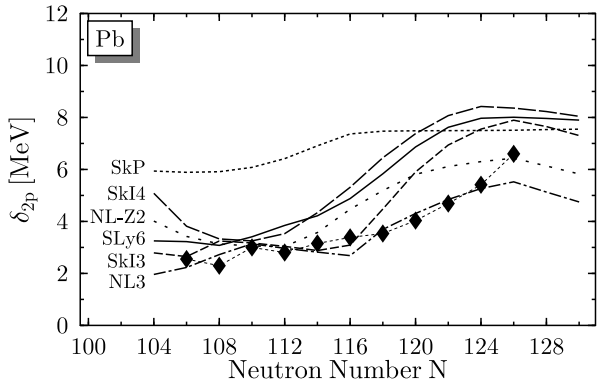}}
\caption{\label{fig:gap:def}
Two-proton gap parameter $\delta_{2p}$ (\protect\ref{eq:tpgap})
for the chain of Pb isotopes
obtained in several  deformed  mean-field models.
Experimental values are marked with filled diamonds.
}
\end{figure}
%
%=========
%
Experimentally, the  even-even $^{176-190}$Hg isotopes have 
oblate ground states but also prolate excited states \cite{Hel99}.
The  data on excitation spectra and charge radii for the Pb isotopes, on
the other hand, are consistent with the assumption that the ground
states down to \mbox{$N=90$} are spherical \cite{Jul01,Jen00,Hey96a}, 
although one has to be aware that the potential landscape for Pb also 
becomes softer for proton-rich isotopes.  As a result, coexisting oblate 
and prolate minima show up \cite{Woo92,Naz93a}, the most spectacular being
the multitude of coexisting structures in  $^{186}$Pb \cite{And00a}. 
A systematic analysis of available excitation data in
Po isotopes demonstrates that collectivity increases rapidly in
neutron-deficient isotopes \cite{Jul01,You97a}.

When deformation effects are taken into account, 
the behavior of  $\delta_{2p}$ changes dramatically. As shown in
fig.~\ref{fig:gap:def}, all our models  predict gradually  decreasing 
$\delta_{2p}$ towards the proton drip line in accordance with experiment.
The crucial point is that one deals  with systems where the mean field 
is  extremely sensitive to small changes in nucleon number. These
rearrangement effects strongly affect the behavior of $\delta_{2p}$
and  mask the  presence of the magic \mbox{$Z=82$} proton shell;
it is  nuclear deformation that explains the apparent discrepancy 
seen in fig.~\ref{fig:gap:sph}.

Complementary information on shell effects can be drawn from $\alpha$-decay
studies. An analysis of reduced $\alpha$ widths and $Q_\alpha$ values of 
nuclei around $^{182}$Pb suggests a weakening of the \mbox{$Z=82$} shell 
effect \cite{Tot99a}. However, since $Q_\alpha$ values are obtained from 
a finite-difference formula similar to $\delta_{2p}$, experimental 
systematics $Q_\alpha$  reflect deformation effects and the structural 
change of ground states between the isotones of Hg, Pb, and Po. Moreover, 
$\alpha$-decay hindrance factor systematics can be  understood assuming 
the stability of the spherical \mbox{$Z=82$} shell at the very proton-rich 
side \cite{Wau94a,Bij95a} and can be described in the language of  
shape mixing \cite{Wau94a,Ric97}.

The rich structure of potential landscapes in the region of proton-rich Pb 
nuclei provides a challenging benchmark for theoretical modeling. 
The quantitative description of all details is beyond the abilities of current 
self-consistent mean-field models and, because of the presence of low-lying
intruder states,  also requires the use of sophisticated techniques which
incorporate configuration mixing effects. Small shifts in single-particle 
levels or changes in the parameterization of  pairing interaction do influence
relative positions of coexisting  minima. Therefore,  many forces miss the 
exact location of the onset of deformation for the Hg isotopes, see 
fig.~\ref{fig:def} and \cite{procAPAC}. Some forces (RMF  and SkI3) even 
predict deformed ground states in   some Pb isotopes -- in contradiction 
with experiment. We have seen in fig.~\ref{fig:def} that the inclusion of 
deformation improves the agreement of theory with experiment for 
$\delta_{2p}$. Still, some  discrepancies remain. Note that all 
SHF forces already overestimate $\delta_{2p}$ for the
doubly magic 
$^{208}$Pb, although this nucleus is spherical, as well as its neighboring
nucleus. The simplest excuse is that the single-particle states in 
$^{208}$Pb are   not perfectly described by the models (see, e.g., 
\cite{Ben99a}). However, another important source of discrepancy are
correlations beyond  the mean field. Their influence on the
two-\emph{neutron} separation  energies around $^{208}$Pb  has been 
studied using the generator  coordinate method (GCM) in
ref.~\cite{Hee01a}. The important outcome is that the quadrupole
correlations decrease the two-neutron shell gap at the shell closure. A
similar effect can be expected for the two-proton shell gap parameter.
The effect  of  quadrupole correlations on $\delta_{2p}$ is probably
largest in the  transitional region of the onset of static deformation
around \mbox{$N=120$} where also the discrepancy between experiment and
our mean-field calculations is largest. Configuration mixing effects in
proton-rich Pb  isotopes were  studied in ref.~\cite{Mey95a}, but were
aimed at excitation  energies, and not mass systematics (see also recent
GCM work
\cite{Cha01}).
%
%=======================================================================
%
\section{Summary}
Finite-difference binding-energy indicators such as  $\delta_{2q}$ or 
$Q_\alpha$ lose their validity as signatures of shell closures as soon 
as they are used in a region where the structure of nuclear ground states 
is rapidly changing. Such  changes of nuclear ground state configurations  
with $Z$ and $N$  explain qualitatively the systematics of experimental 
$\delta_{2p}$ and $Q_\alpha$ in the neutron-deficient Pb region. In 
particular, our calculations do not support previous speculations
about quenching of the \mbox{$Z=82$} shell in the neutron-deficient 
Pb isotopes.

It is important, however, to remember that the complex structure of 
nuclei from shape-coexisting  regions limits their description in 
terms of mean-field models. The presence of low-lying intruder states
having  different shapes requires the use of theories which can account for 
configuration mixing effects. Therefore, the whole notion of a ``shell gap", 
which originates from the mean-field theory, can be questionable in such 
situations. In general, one has to be aware that the usual signature of 
spherical shell closures might not be robust  when going to exotic nuclei
far from stability in which correlations play an important role.
%
%=======================================================================
%
\section*{Acknowledgements}
Discussions  with H.~Geissel, H.~Grawe, G.~M{\"u}n\-zen\-berg, 
Ch.~Schei\-den\-berger, and H.~Wollnik are gratefully acknowledged. 
This work was supported in part by
Bundesmi\-ni\-ste\-rium f\"ur Bildung und Forschung (BMBF), Project
No.\ 06 ER 808, by Gesellschaft f\"ur Schwer\-ionen\-for\-schung (GSI),
and by  the U.S.\ Department of Energy under Contract Nos.\ DE-FG02-96ER40963
(University of Tennessee), DE-FG05-87ER40361 (Joint Institute for Heavy
Ion Research), and DE-AC05-00OR22725 with UT-Battelle, LLC (Oak Ridge 
National Laboratory).
%
%=======================================================================
%
 
%
%=======================================================================
%
\end{document}